\documentclass{emulateapj}
\lefthead{GAUDI} \righthead{{\it KEPLER} AND KUIPER}
\def\kms{{\rm km\,s^{-1}}}
\def\eq#1{equation (\ref{#1})}

\def\km{\rm km}
\def\au{\rm AU}
\def\kms{\rm km~s^{-1}}

\def\thk{\theta_{\rm K}}
\def\thf{\theta_{\rm F}}
\def\ths{\theta_*}
\def\rhof{\rho_{\rm F}}
\def\rhos{\rho_{\rm *}}
\def\kep{{\it Kepler}}
\def\deg{^\circ}
\def\rb{r_{\rm b}}
\def\mb{m_{\rm b}}
\def\muas{\mu{\rm as}}
\def\atc{\left<t_c\right>}
\def\qmin{Q_{\rm min}}
\def\texp{t_{\rm exp}}
\def\ndet{N_{\rm det}}

\begin{document}

\title{{\it Kepler} and the Kuiper Belt}
\author 
{B.\ Scott Gaudi}
\affil{Harvard-Smithsonian Center for Astrophysics, 60 Garden St., Cambridge, MA 02138}

\begin{abstract}
The proposed field-of-view of the {\it Kepler} mission is at an
ecliptic latitude of $\sim55^\circ$, where the surface density of
scattered Kuiper Belt Objects (KBOs) is a few percent that in the
ecliptic plane.  The rate of occultations of \kep\ target stars by
scattered KBOs with radii $r\ga10~\km$ is $\sim 10^{-6}-10^{-4}~{\rm
star^{-1}}~{\rm yr^{-1}}$, where the uncertainty reflects the current
ignorance of the thickness of the scattered KBO disk and the faint-end
slope of their magnitude distribution.  These occultation events will
last only $\sim 0.1\%$ of the planned $\texp=15$ minute integration
time, and thus will appear as single data points that deviate by tiny
amounts.  However, given the target photometric accuracy of {\it
Kepler}, these deviations will nevertheless be highly significant,
with typical signal-to-noise ratios of $\sim 10$.  I estimate that
$\sim 1-20$ of the $10^5$ main-sequence stars in {\it Kepler}'s
field-of-view will exhibit detectable occultations during its
four-year mission.  For unresolved events, 
the signal-to-noise of individual occultations
scales as $\texp^{-1/2}$, and the minimum detectable
radius could be
decreased by an order of magnitude to $\sim 1~\km$ by
searching the individual $3~{\rm sec}$ readouts for occultations.  
I propose a
number of methods by which occultation events may be differentiated
from systematic effects. {\it Kepler} should measure or significantly
constrain the frequency of highly-inclined, $\sim 10~\km$-sized KBOs.
\end{abstract}
\section{Introduction\label{sec:intro}}

The Kuiper belt (see
\citealt{lj02} for a review) constitutes one of the few fossil records of the
formation of planets and planetesimals in the early solar system.  It
is therefore of great interest to determine the ensemble properties of
its denizens.  These properties provide clues to the physical
processes operating in the protoplanetary solar disk, as well as in
the subsequent evolution of the solar system.

Kuiper Belt Objects (KBOs) are currently discovered in deep optical
imaging surveys that are primarily limited to latitudes within $\sim
10\deg$ of the ecliptic or invariable plane (see, e.g., \citealt{millis02}).  
The apparent $R$-magnitude of
a KBO at a distance of $a\simeq 40~\au$ with radius $r \simeq 100~\km$
is $m_R\simeq 23.4$, assuming an albedo of $4\%$.  Since the flux of a
KBO arises from reflected sunlight, it is proportional to $r^{2}$ and
$a^{-4}$.  Therefore, imaging surveys are only able to detect relatively large
and nearby KBOs with reasonable expenditure of resources.
Furthermore, in order to have the best chance of success, they do
not target high ecliptic latitudes, and are therefore biased against
KBOs with large inclinations $i$.

Large KBOs are known to fall into at least three dynamical classes.
Classical KBOs are relatively dynamically cold, with eccentricities
$e\la 0.3$ and $i\la 5\deg$.  Resonant KBOs are in mean
motion resonances with Neptune, and were likely trapped there
and subsequently excited to high $i$ and $e$ by the outward migration of
Neptune \citep{malhotra95}.  Scattered KBOs have very broad $e$ and $i$
distributions, and have probably been excited by some as-yet unknown process.

The size distribution of large KBOs is known to follow the form of
$r^{-q}$, with $q\simeq 4$, and a surface density of objects of
$1~{\rm deg^{-2}}$ at $r\simeq 100~\km$.  For $r\la 100~\km$,
collisions will preferentially destroy KBOs, so that below a certain
radius $r<\rb\la 100~\km$, the size distribution is expected to
flatten.  This break radius is set by the dynamical and physical
properties of the KBOs, and is roughly where the collision time is
equal to the age of the system.  The first estimate of this break
radius, based on deep Hubble Space Telescope (HST) observations, yields
$\rb \sim 70~\km$ (\citealt{bern04}, hereafter B04).

Unfortunately, it will be difficult, in the near future, to learn
about the properties of small KBOs beyond what is already known using
direct imaging, and especially so for those with high inclinations.
Further progress will require next-generation synoptic surveys
(Pan-STARRS, \citealt{kaiser02}; The Discovery Channel Telescope, \citealt{mds03}; LSST, \citealt{tyson02}), or new methods.  One such method,
first proposed by \citet{bailey76}, is to search for stellar
occultations by foreground KBOs.  This technique has been explored by
several authors \citep{dyson92,bw97,cf04}, and is being implemented by
\citet{roques03}, and
the Taiwanese-American Occultation Survey (TAOS; \citealt{liang03}).
However, due to the low expected event rate, the first occultation
surveys will likely focus on low-latitude fields.  Therefore, the
statistics of small, high-inclination KBOs will remain meager.

Here I explore the possibility of measuring the surface density of
small KBOs with high inclinations using the NASA space mission \kep.
\kep's main goal is to detect transiting planets around normal stars,
and it is far from ideally suited to detect occultations by foreground
KBOs, especially given the high latitude ($55\deg$) of its target
field, and its long ($15~{\rm min}$) exposure time.  
However, due to the large number of target stars and
extraordinary photometric precision, I demonstrate that it may
nonetheless detect a significant number of occultation events.  If
these can be separated from systematic artifacts, {\it Kepler} should
be able to determine or severely constrain the frequency of small,
high-inclination KBOs.

\clearpage

\section{Order-of-Magnitude Estimates\label{sec:oom}}

I begin by providing order-of-magnitude estimates of the rate and
properties of KBO occultation events in \kep's field-of-view (FOV).
Assuming all KBOs are at the same distance, the optical depth to occultations is roughly the differential
angular surface density $\Sigma_r(r)={\rm d}N(r)/{\rm d}r{\rm d}\Omega$
of KBOs of radius $r$ in the \kep\ FOV, times the solid angle
$\Omega(r)$ subtended by a KBO of radius $r$, integrated over all
radii,
\begin{equation}
\tau \sim \int {\rm d}r \Omega(r) \Sigma_r(r).
\label{eqn:tausimple}
\end{equation}
Although the optical depth to KBOs at a constant distance is most
directly related to their size distribution, optical surveys do not
directly constrain this quantity.  Rather, they constrain
the distribution of apparent magnitudes.  I therefore switch
the integration variable in \eq{eqn:tausimple} to apparent 
$R$-magnitude $m_R$,
\begin{equation}
\tau \sim \int {\rm d}m_R \Omega(m_R) \Sigma(m_R).
\label{eqn:tausimplem}
\end{equation}
Assuming constant albedo and
distance, the radius of a KBO is related to its $R$-magnitude by
\begin{equation}
r(m_R)=r_{23}10^{-0.2(m_R-23)}.
\label{eqn:radmag}
\end{equation}
where $r_{23}$ is the radius of a KBO at $m_R=23$. 
The measured differential surface density $\Sigma(r)={\rm d}N(r)/{\rm d}m_R{\rm d}\Omega$ of all bright KBOs as a function of
apparent magnitude near the ecliptic has the form
$\Sigma(m_R)=\Sigma_{23}10^{\alpha(m_R-23)}$, where
$\alpha$ can be related to the power-law index of the
size distribution by $q=5\alpha+1$.  Assuming that the
latitude distribution of scattered KBOs is a Gaussian with a
standard deviation $\sigma_\beta$, and that a fraction $f_{\rm SK}$ of KBOs
detected near the ecliptic plane are scattered KBOs,
\begin{equation}
\tau  \sim f_{\rm SK} \pi \theta_{23}^2 \Sigma_{23}
e^{\frac{-\beta^2}{2\sigma_\beta^2}}\int {\rm d}m_R
10^{(\alpha-0.4)(m_R-23)}.
\label{eqn:tau1}
\end{equation}
Here $\theta_{23}\equiv r_{23}/a$ is the angular size of a KBO with
$m_R=23$.

Assuming that objects at the break magnitude
produce occultations above the detection limit, 
there are two regimes to consider.  If the slope of the magnitude
distribution above the break magnitude $m_R>\mb$ ($r<\rb$) is
$\alpha_2 \la 0.4$, then the integral in \eq{eqn:tau1} is dominated by
objects near $m_b$, and is 
$\sim \Delta m_R 10^{(\alpha-0.4)(m_b-23)}$, where $\Delta m_R\simeq 
0.6|\alpha-0.4|^{-1}\simeq 3$ is roughly the full-width half-max of the
distribution.  In the other regime, if $\alpha_2\ge 0.4$, then the
integral in \eq{eqn:tau1} formally diverges.  In practice, the
lower limit in radius, or 
upper limit in magnitude, $m_{\rm max}$, will then be set by
where the occultation is too brief or too shallow to be detectable.
As I demonstrate in \S\ref{sec:rate}, the former generally turns out to be satisfied
first. 
In this regime, the integral in \eq{eqn:tau1} is dominated by
objects with $m_R\sim m_{\rm max}$, and thus is $\sim \Delta m_R
10^{(\alpha-\alpha_2)(\mb-23)+(\alpha_2-0.4)(m_{\rm max}-23)}$. 

A simple boxcar occultation of duration $2t_c$ has a 
signal-to-noise $Q$ of,
\begin{equation}
Q={2\atc \gamma^{1/2}}{t_{\rm exp}^{-1/2}},
\label{eqn:sstn}
\end{equation}
where $\gamma=2.46 \times 10^{5}~{\rm s^{-1}}$ is the rate at which
photons will be collected by \kep\ from a star with $m_*=12$.  This
expression holds when the occultation duration is less than the
exposure time.  For circular KBOs and equatorial occultations, 
the duration is simply $t_c=\thk/\mu$, 
where $\thk$ is the angular
radius of the KBO, $\mu=v/a$ is the proper motion of the KBO,
and $v\sim v_\oplus =30~\kms$ is the relative velocity of the KBO,
which is primarily due to the reflex motion of the
Earth.  This yields $t_c \simeq 4~{\rm s}~10^{-0.2(m_R-23)}$.
For objects near the break of $m_b=26$,
$t_c\simeq 1~{\rm s}$, and for the planned \kep\ exposure time of $\texp=15~{\rm min}$, a typical target star with apparent magnitude $m_*=14$ yields
$Q\sim 13$.  The minimum acceptable signal-to-noise of $Q\sim 7$
(justified in \S\ref{sec:rate}) is reached at $m_{\rm max}\simeq 27.4$, 
where $r\simeq 16~{\rm km}$, and $t_c\simeq 0.5~{\rm s}$.

Adopting fiducial values for the various input parameters, I now
estimate the optical depth in the two regimes.  Surveys of bright KBOs
find $\alpha\simeq 0.6$ and $\Sigma_{23}\sim 1~{\rm
deg^{-1}~mag^{-1}}$ (e.g., \citealt{lj02}).  The properties of 
the scattered KBO
population are poorly known due to severe selection biases, and
estimates of these properties should therefore be taken with caution.
I adopt $\sigma_i =20\deg$ (\citealt{brown01}, hereafter B01; 
\citealt{tjl01}), $f_{\rm SK}\sim 0.1$, and 
a common distance of $a=42~\au$.  Assuming a
4\% albedo, this yields $r_{23}\simeq 123~\km$, and 
$\theta_{23}\simeq 4~{\rm mas}$.   For classical KBOs, the
break radius is expected to occur near $m_b\sim 24$ \citep{ps04}, as is
observed (B04).  However, scattered KBOs may have longer collision
times, and thus the break radius may be smaller.  I adopt
$m_b\simeq 26$, as B04 find for their `excited' subsample.
Finally, the FOV of \kep\ is at $\beta\simeq 55\deg$. Altogether, this yields
$\tau_{\rm br} \sim 10^{-13}$
for the break-limited regime, and
$\tau_{\rm sn} \sim 3 \times 10^{-13}10^{1.4\alpha_2}$
for the signal-to-noise limited regime.

The event rate $\Gamma$ is related to the optical depth by,
\begin{equation}
\Gamma = \frac{2\tau}{\pi\atc},
\label{eqn:erate}
\end{equation}
where $\atc$ is the mean time scale of the occultation events. 
Assuming that $\atc$ is equal to the time scale at the peak of
the size distribution, this yields $\Gamma_{\rm br} \simeq 2\times 10^{-6}~{\rm yr^{-1}}$
for the break-limited regime, and
$\Gamma_{\rm sn} \simeq 10^{-6}~{\rm yr^{-1}} 10^{1.4\alpha_2}$
for the signal-to-noise limited regime.

\kep\ will have
$N_*\sim 10^5$ main-sequence stars in its field-of-view, and will have
a lifetime of $T=4$ years, so the number of expected events is $N_{\rm
det}=N_*\Gamma_{\rm br} T \simeq 0.8$ for the break-limited regime.  
For the signal-to-noise limited regime, and 
a faint-end slope of $\alpha_2=0.6$, $\Gamma_{\rm sn}\simeq 7.7\times 10^{-6}~{\rm yr^{-1}}$, and the number of expected events is 
$N_{\rm det}=N_*\Gamma_{\rm sn} T \simeq 3$.

\section{Evaluation of Event Rate}\label{sec:rate}

I now provide a somewhat more accurate estimation of the occultation
rate in \kep's FOV, taking into account the distribution of KBO radii,
inclinations, and velocities; the number, magnitude, and angular size
distribution of the target stars; and the suppression of
signal-to-noise in occultation events due to diffraction and finite
source size effects.  Readers who are not interested in these details
can skip to \S\ref{sec:discussion} without significant loss of continuity.
I will continue to make a number of simplifying
assumptions, including that all KBOs are located at the same distance
of $a=42~\au$.  

The observable optical depth to a star of a given magnitude $m_*$ and
angular size $\ths$ is
\begin{equation}
\tau(m_*,\ths)=\int {\rm d}m_{R} \Sigma(\beta,m_{R})\Omega(m_{R})
\Theta(Q-\qmin),
\label{eqn:tauacc}
\end{equation}
where $\beta$ is the ecliptic latitude of the FOV, $\qmin$ is the
minimum detectable signal-to-noise, and $\Theta(x)$ is the Heaviside
step function.

The surface density $\Sigma(\beta,m_{R})$ of KBOs of a given $m_R$ at
an ecliptic latitude $\beta$ can be written as,
$\Sigma(\beta,m_{R})=\Sigma(0,m_{R})f_\Sigma(\beta)$, where
$f_\Sigma(\beta)$ is the surface density distribution of KBOs
normalized to the ecliptic plane.  For a population with an observed
inclination distribution at the ecliptic plane of $f_e(i)$, this is
given by (B01),
\begin{equation}
f_\Sigma(\beta)= \int_\beta^{\pi/2} {\rm d}i \frac{f_e(i)\sin
i}{(\sin^2i-\sin^2\beta)^{1/2}}\times\left[ \int_0^{\pi/2}{\rm d}i
f_e(i)\right]^{-1}.
\label{eqn:ldist}
\end{equation}
An inclination distribution of the form
$f_e(i)=\exp(-i^2/2\sigma_i^2)$, with a standard deviation $\sigma_i=20\deg
\pm 4\deg$, provides a reasonable fit to the observed distribution of
scattered KBOs (B01; \citealt{tjl01,millis02}). For this form, the denominator of \eq{eqn:ldist} is
equal to $(\pi/2)^{1/2}\sigma_i {\rm Erf}(\pi/2\sqrt{2}\sigma_i)$.  The
\kep\ FOV extends over $10\deg$ in ecliptic latitude.  Since
the surface density varies by a factor of $\sim 2$ over this range
for reasonable values of $\sigma_i$, I average $f_\Sigma(\beta)$  over $50\deg\le \beta \le
60\deg$, assuming a uniform distribution in $\beta$.

For unresolved events ($2t_c \le \texp$), I write the 
signal-to-noise $Q$ of an occultation as,
\begin{equation}
Q={2 t_c\gamma^{1/2} }{t_{\rm
exp}^{-1/2}}f_{u_0}f_{Q}10^{-0.2(m_*-12)},
\label{eqn:snexp}
\end{equation}
whereas, for resolved events ($2t_c \ge \texp$),
\begin{equation}
Q=(2t_c\gamma)^{1/2} f_{u_0}f_{Q} 10^{-0.2(m_*-12)},
\label{eqn:snexpres}
\end{equation}
Where, as before, $t_{\rm exp}$ is the exposure time and
$\gamma$ is the photon collection rate at $m_*=12$.  The term 
$f_{u_0}$ is the reduction in signal-to-noise due to
non-equatorial occultations, which is
$f_{u_0}=\int_0^1 {\rm d}u_0(1-u_0^2)^{1/2}=0.79$ for unresolved events,
and $f_{u_0}=\int_0^1 {\rm d}u_0(1-u_0^2)^{1/4}=0.87$ for resolved
events.

Significant
suppression of the signal-to-noise due to diffraction and finite-source
effects is expected 
whenever the angular radius $\thk$ of the KBO is
comparable to, or smaller than, either the angular size of the star
$\ths$, or the Fresnel angle $\thf=\sqrt{\lambda/2\pi a}$, where
$\lambda$ is the wavelength of observations.  
I determine $f_Q$ by numerically calculating occultation light curves
for various values of $\rhof\equiv \thf/\thk$ and $\rhos\equiv\ths/\thk$, integrating over a bandpass
centered at $\lambda=500~{\rm nm}$ with a full width of $25\%$.  I
then determine the suppression in signal-to-noise of these curves relative to a
boxcar complete occultation with duration $2t_c$.  
For unresolved events, for which diffraction and finite-source effects
may be important, this can be approximated as
\begin{equation}
f_Q=\frac{1}{4}
{\rm Erfc}\left(\frac{\log{\rhos}-0.52}{0.42}\right)
{\rm Erfc}\left(\frac{\log{\rhof}}{0.65}\right).
\label{eqn:fq}
\end{equation}
For a mid-F dwarf at $\sim 1~{\rm kpc}$ (typical of Kepler target
stars), $\ths \simeq 6~\muas$, while for $\lambda =500~{\rm nm}$ and
$a=42~\au$, $\thf\simeq 23~\muas$.  Therefore the diffraction effects
are typically expected to be more important than the finite size of
the stars.  The radius where $\thk=\thf$ is $r \sim 1~\km$, or $m_R
\sim 34$; the radius where $\thk=\ths$ for the example above is $r\sim
0.26~{\rm km}$, or $m_R \sim 37$.  As both of these magnitudes are
considerably fainter than the expected
cutoff in $Q$ due to photon noise for the nominal \kep\ exposure time,
the corrections due to diffraction and finite source effects are
small in this case.

The final ingredient in calculating $Q$ is the radius crossing
time $t_c$.  I determine the relative velocity $v$ of a KBO at a given
ecliptic longitude and latitude by $v=|{\bf v}_{\perp,\oplus}-{\bf
v}_K|$, where ${\bf v}_{\perp,\oplus} ={\bf v}_\oplus- ({\bf v_\oplus}
\cdot {\hat {\bf n}}) {\hat {\bf n}}$, ${\hat {\bf n}}$ is the unit
line-of-sight vector to the KBO, and $v_K\equiv v_\oplus(a/\au)^{-1/2}$.  I assume circular orbits for the
Earth and KBO, and that the KBO is observed at a latitude equal to its
inclination.  I then average the proper motion $\mu=v/a$ over
ecliptic longitude.  This yields $\left<\mu\right>=3.19''~{\rm
hr^{-1}}$, and $t_c=3.70~{\rm s}$ for $r=100~\km$.

The average radius crossing time $\atc$ is given by,
\begin{equation}
\left<t_c(m_*,\ths)\right>=\int {\rm d}m_{R} \Sigma(\beta,m_{R})
\frac{\thk(m_{R})}{\left<\mu\right>} \Theta(Q-\qmin)
\label{eqn:atc}
\end{equation}
$$
\times \left[\int {\rm d}m_{R} \Sigma(\beta,m_{R}) \Theta(Q-\qmin)\right]^{-1}.
$$
This can be combined with the optical depth $\tau(m_*,\ths)$ to yield
the event rate,
\begin{equation}
\Gamma(m_*,\ths)=\frac{2\tau(m_*,\ths)}{\pi\left<t_c(m_*,\ths)\right>}.
\label{eqn:erateacc}
\end{equation}
Finally, the total number of detected occultations can be found by
integrating over the magnitude distribution ${\rm d}N_*/{\rm d}m_*$
and angular size distribution ${\rm d}N_*/{\rm d}\ths$ of the stars in
\kep's FOV,
\begin{equation}
\ndet= T \int {\rm d}m_* \frac{{\rm d}N_*}{{\rm d}m_*} \int {\rm
d}\ths \frac{{\rm d}N_*}{{\rm d}\ths} \Gamma(m_*,\ths).
\label{eqn:rate}
\end{equation}
I adopt the joint apparent magnitude -- spectral type distribution
derived by \citet{jd03} and presented in their Table 1; this
distribution is in turn based on the Besan\c{c}on Galactic model
\citep{robin03}, assuming an extinction of $1~{\rm mag~kpc^{-1}}$.
Adopting the relations between absolute magnitude, physical radii, and
spectral type for main-sequence sequence stars from \citet{allen76},
the corresponding distribution of angular radii can also be derived.

\begin{figure}
\epsscale{1.7} 
\plotone{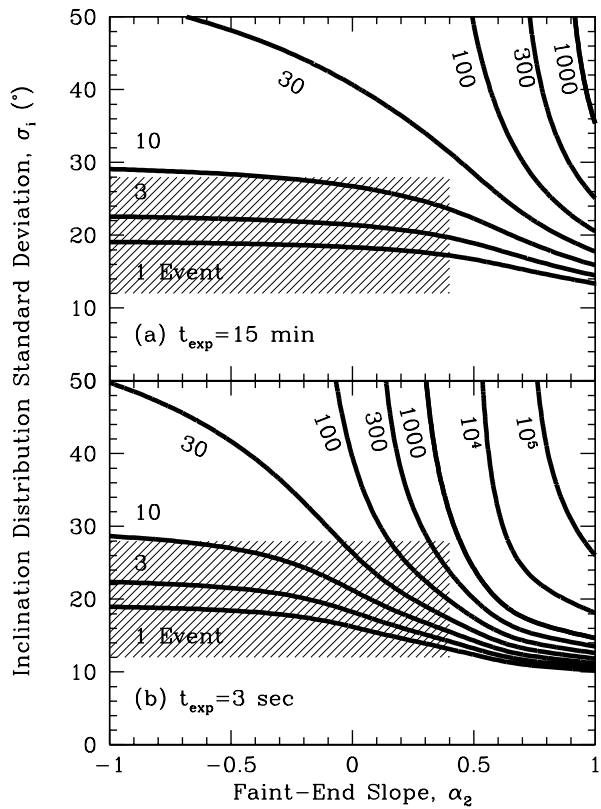}
\vskip-0.3in
\caption{\label{fig:one} 
The contours show the number of expected occultation events
with signal-to-noise $>7$ in \kep's field-of-view, as a function of
the standard deviation $\sigma_i$ of the inclination distribution in degrees,
and the faint-end slope $\alpha_2$ of the surface density versus
magnitude relation.  The shaded region shows the approximate $95\%$
confidence limits on these parameters.  Each panel shows the number of events
for a different exposure time.  (a) Nominal \kep\ exposure time of
$\texp=15\min$. (b) Exposure time equal to the readout time
of 3 seconds.}
\end{figure}

For the KBO magnitude distribution $\Sigma(0,m_{R})$ near the ecliptic
plane, I use the results from B04.  They compile data from several
surveys, including their own HST results, and
fit the surface density distribution to a double power-law form,
\begin{equation}
\Sigma(m_R)=(1+c)\Sigma_{23}\left[10^{-\alpha_1(m_R-23)}+c10^{-\alpha_2(m_R-23)}\right]^{-1},
\label{eqn:sigmadouble}
\end{equation}
where $c\equiv 10^{(\alpha_1-\alpha_2)(m_b-23)}$.
For their `excited' subsample, which has
$i>5\deg$, and $d<28\au$ or $d>55\au$, they find $\alpha_1=0.66$, $m_b=26.0$, and
$\Sigma_{23}=0.39~{\rm mag^{-1}~deg^{-2}}$.  I estimate
that $\sim 1/3$ of these are scattered KBOs based on the
relative number of scattered to resonant KBOs in the B01 analysis.
The faint-end slope
$\alpha_2$ is poorly constrained for this sample, but is bounded at
$68\%$c.l.\ by $\alpha_2\la 0.1$ and $95\%$c.l.\ by $\alpha_2<0.4$.

To calculate the rate of detected events, I adopt a signal-to-noise
threshold of $\qmin=7$.  The total number of flux measurements made
by \kep\ over its lifetime will be $\sim N_* T
t_{\rm exp}^{-1} \sim 4\times 10^9$, therefore requiring a false alarm
probability of $\la 10^{-10}$.  One expects $\la 1$ false alarm
for $Q>7$.  At the break magnitude $m_{\rm eq}$, the
signal-to-noise is $Q\ga 10$ for $\texp=15~{\rm min}$.  
Therefore, the results are not
sensitive to the choice of $\qmin$ for $\qmin \la 10$, as long as
objects near the break magnitude dominate the integral of the
magnitude distribution, i.e.\ as long as $\alpha_2\la 0.4$.

Figure \ref{fig:one}(a) shows number of detected occultation events in
\kep's FOV, as a function of the dispersion in the inclination
distribution $\sigma_i$, and the faint-end slope $\alpha_2$.  The
change in the shape of the contours near $\alpha_2\sim 0.4$ is due to
the transition from the break limited to signal-to-noise limited
regimes described in the previous section.  For $\alpha_2\la 0.4$, the
event rate depends mainly on $\sigma_i$, whereas for $\alpha_2\ga 0.4$, it
depends primarily on $\alpha_2$.  Very few events are expected for
$\sigma_i\la 10\deg$ regardless of $\alpha_2$.  This is due to the
exponential suppression of the number of KBOs in the \kep\ FOV for the
assumed Gaussian distribution of inclinations.  For the fiducial value
of $\sigma_i=20\deg$, and $-1.0 \la \alpha_2\la 0.2$, the number of events is $\sim 2$.
This is a factor of a few larger than the rough estimate in the
previous section for the break-dominated regime.  This difference
can be attributed to differences in the assumed number of target stars, 
surface density at the ecliptic, and suppression at the latitude
of the \kep\ FOV.  The event rate
increases steeply for increasing $\alpha_2\ga 0.4$.  For $\alpha_2=0.6$,
$\ndet\simeq 8~{\rm yr^{-1}}$, again a factor of a few larger than
the rough estimate.

\section{Discussion}\label{sec:discussion}

The shaded area in Figure \ref{fig:one} shows roughly the 95\%
confidence limit on the standard deviation $\sigma_i$ of the inclination
distribution from B01, and the faint-end slope $\alpha_2$
from B04.  In this region, the predicted number of events
varies from $\ll 1$ to nearly $20$.  Marginalizing $\ndet$ over
$\sigma_i$ and $\alpha_2$, assuming Gaussian uncertainty distributions
in both parameters with mean and dispersion of $20\deg$ and $4\deg$ for
$\sigma_i$, and $-0.5$ and $0.6$ for $\alpha_2$, yields a best
estimate of the number of expected events of $\left< \ndet \right>\sim 3$.  If no events
are detected during the mission lifetime, a 95\% confidence upper
limit can be placed that should correspond to the $\ndet=3$ contour,
which would exclude roughly half the currently allowed region of
$\sigma_i$ and $\alpha_2$.

Although the expected number of events in the 95\% c.l.\ region of
$\sigma_i$ and $\alpha_2$ is relatively small, it is important to keep
in mind that these parameters might be considerably in error, as all
previous surveys have had relatively strong selection biases against
scattered KBOs.  In particular, the constraint on $\alpha_2$ from B04
is applicable to their `excited' sample, the majority of which are
probably resonant KBOs.  
Therefore, the faint-end slope of scattered KBOs may differ
considerably from that measured by B04; this might even be expected if
scattered KBOs were excited early in the formation of the solar
system.  For $\alpha_2\ga 0.4$, the number of expected
events can be quite substantial.  In addition
to uncertainties in the properties of the known scattered KBO population, there may be additional populations of KBOs at larger distances that are
currently undetectable, but may be detected or constrained by \kep.

For unresolved events, the signal-to-noise of occultations scales as $\texp^{-1/2}$.  Therefore, the signal-to-noise of individual events
can be increased by simply decreasing the \kep\ exposure time.  
In the break-limited regime ($\alpha_2 \la 0.4$), where small
objects with $m_R>m_b$ do not contribute much to the optical depth, this does not
significantly increase 
the number of detectable events, although it does improve the reliability
of already detectable occultations with $m_R\la m_b$.
In the signal-to-noise limited regime $\alpha_2 \ga 0.4$, decreasing
the exposure time can increase the number of detectable occultations
dramatically.   As currently planned, \kep\ will only transmit 
integrations of $\texp=15~{\rm min}$, due primarily to bandwidth limitations. 
However, the detectors will in fact be read out every 3 seconds in 
order to avoid saturation by bright stars.  These $3~{\rm sec}$ readouts
will then be filtered and combined into $15~{\rm min}$ integrations.  
It may be possible, using in-flight software, to search for occultations in 
the individual $3~{\rm sec}$ exposures, and flag and transmit 
significant events.   With this possibility in mind, in 
Figure \ref{fig:one}(b) I show $\ndet$ as a function
of $\sigma_i$ and $\alpha_2$ for $\texp=3~{\rm sec}$.  As expected,
deep in the break-limited regime ($\alpha_2\la 0$), 
the number of detectable occultations is similar 
for $\texp=3~{\rm sec}$ and $\texp=15~{\rm min}$.
However, for $\alpha_2\ga 0$, the number of expected event increases
significantly for $\texp=3~{\rm sec}$.  In the
allowed region of parameter space, the number of events can be as large as $\ndet\sim
600$.  Marginalizing over $\sigma_i$ and $\alpha_2$ yields $\left<
\ndet \right> \sim 12$.  For $\alpha_2=0.6$, and $\sigma_i =20\deg$,
$\ndet \simeq 10^3$.  Whether or not it is efficient or
feasible to search the individual $3~{\rm sec}$ integrations for
occultations remains to be seen. However, given the large potential 
increase in the
number of expected events, and the fact that the subsequent constraints
will be hard to acquire with other methods, the possibility should be
considered.

Given the fact that the occultation events are unresolved, and have a
small amplitude, systematic effects that could mimic occultation
events are a serious concern.  For a large sample of candidate events,
there are some signatures that may differentiate between real
occultations and systematics.  Occultation events should show a factor
of $\sim 2$ gradient in the optical depth over the $\sim 10\deg$ in
ecliptic latitude spanned by the \kep\ FOV.  In addition, the
relative velocity also varies over the course of the year; for the
latitude of the \kep\ FOV, this variation is $10~\kms$ peak-to-peak,
which corresponds to a variation in both the signal-to-noise and event
rate of amplitude $\sim 30\%$.  Unfortunately, given the intrinsic
velocity dispersion of KBOs, and the variation in signal-to-noise
noise due to non-equatorial occultations, it will be difficult to
measure these deviations unless there are a large number of events.
For individual events, it may be possible to distinguish between
occultations and systematic effects based on the detailed shape and
centroid of the stellar point-spread function for high signal-to-noise events
 \citep{jd03}.
Ground-based follow-up observations can be used to eliminate
contamination from background stars that happen to be blended with the
primary star.  

It may also be possible to directly detect the KBOs causing 
occultation events by imaging them in reflected light. 
Provided that the data are analyzed
in real time, candidate occultation events could be alerted within a day
of the event.  The signal-to-noise of the occultations should be
correlated with their size (e.g., Eq.\ \ref{eqn:snexp}), and therefore
brighter events can be preferentially selected for follow-up.  The
brighter candidates will have $m_R \sim m_B \simeq 26~{\rm mag}$,
within the range of large, ground-based telescopes.  Typical proper
motions are a few arcseconds per hour.  Therefore, deep imaging taken
at several epochs over a few days after the occultation should reveal
a faint object moving away from the bright target star.  

Provided that systematic effects can be controlled, and a reliable
sample of occultation events can be acquired, then the observed
distribution of signal amplitudes can be used to statistically infer
the distribution of event time scales.  This can then be converted to
an estimate of the optical depth.  If no events are detected, then it
should be possible to place limits on the surface density of KBOs in
the \kep\ FOV.  Therefore, \kep\ should measure or significantly constrain the
number of highly-inclined, $\sim 10~\km$-sized KBOs.

\acknowledgments 
I would like to thank Josh Bloom, Dave Latham, and Josh Winn for useful discussions, Matt Holman for discussions and reading the manuscript, the anonymous referee for helpful comments and suggestions, and Cheongho Han
for the use of his code to calculate occultation light curves. 
This work was supported by a Menzel Fellowship from the 
Harvard College Observatory.

\end{document}